\documentclass[twocolumn]{revtex4}
\usepackage[dvipdfm]{graphicx}
\usepackage{fancybox}
\usepackage{amsmath,amssymb,bm,amsthm}
\def\beq{\begin{equation}}
\def\eeq{\end{equation}}

\def\<{\langle}
\def\>{\rangle}
\def\rt#1{\sqrt{\mathstrut #1}}

\renewcommand{\d}{\partial}

\begin{document}

\title{Asymptotic forms and scaling properties of the relaxation time near threshold points in 
spinodal type dynamical phase transitions}

\author{Takashi Mori}

\author{Seiji Miyashita}
\affiliation{
Department of Physics, Graduate School of Science,
University of Tokyo, Bunkyo-ku, Tokyo 113-0033, Japan
}
\affiliation{
CREST, JST, 4-1-8 Honcho Kawaguchi, Saitama, 332-0012, Japan}
\author{Per Arne Rikvold}
\affiliation{Department of Physics and Center for Materials Research and 
Technology, Florida State University, Tallahassee, Florida 32306-4350, USA}
%%miya0426
%\author{William Klein}
%\affiliation{Department of Physics, Boston University, Boston, MA 02215, USA}

%
\begin{abstract}
We study critical properties of the relaxation time at a threshold point
in switching processes between bistable states under change of external fields. 
%%%
In particular, we investigate the relaxation processes near
the spinodal point of the infinitely long-range interaction 
model (the Husimi-Temperley model) by analyzing the scaling properties of the 
corresponding Fokker-Planck equation. We also confirm the obtained scaling relations
 by direct 
numerical solution of the original master equation, and by kinetic Monte 
Carlo simulation of the stochastic decay process. 
%%%
In particular, we study the asymptotic forms of the divergence of the relaxation time 
near the spinodal point, and reexamine its scaling properties. 
%%%
We further extend the analysis to transient critical phenomena 
such as a threshold behavior with diverging switching time
under a general external driving perturbation. 
This models photo-excitation processes in spin-crossover materials.
In the ongoing development of nano-size fabrication, such size-dependence of 
switching processes should be an important issue, and 
the properties obtained here will be applicable to a wide range of physical  processes.
\end{abstract}

\pacs{75.50.Xx}

% 75.10.Jm(Quantized spin models)
% 75.45.+j(Macroscopic quantum phenomena in magnetic systems)
% 75.75.+a(Magnetic properties of nanostructures)

\maketitle

%*******************************************************************************
%\begin{document}

\section{Introduction}

Relaxation phenomena in strongly interacting systems have been studied 
extensively, including various types of threshold phenomena. 
One of the most typical examples is the decomposition at the coercive field 
(the end of the hysteresis loop of a ferromagnetic system).
This phenomenon appears in the field dependence of the order parameter
in the ordered phase.
In mean-field studies, this phenomenon is well described by the change 
in the free energy as a function of
the magnetization (the order parameter of a ferromagnetic system).
That is, in the ordered state, the free energy has two minima
representing the symmetry-broken states.
When we apply an external field, one of them is selected to be the
equilibrium state.
For a weak field, however, the other state remains as a metastable state.
When the field becomes strong, the metastable state finally becomes unstable.
This point is called the spinodal point. 
However, in models with short-range interactions, fluctuations cause the
system to escape from the metastable state through 
nucleation phenomena, and thus the change of the relaxation time is
smeared.
Thus, although there are several studies on the divergence of the 
relaxation near the spinodal points \cite{Binder1973}, 
the spinodal phenomenon in short-range systems
must be defined only as a crossover \cite{Rikvold1994}.  
%The nature of relaxation depend the size of the system in such short range models. 
%Therefore, finite sizes scaling analysis of the relaxation time near the spinodal point 
%has not studied seriously in a unified way as far as we know.

Recently, however, it has been pointed out that the mean-field
universality class is realized in spin-crossover type
systems \cite{Miyashita2008}.
In these materials, the volume of the unit cell changes, depending on the
local two-fold states (say the high-spin (HS) and low-spin (LS) states).
The volume change causes a lattice distortion, and the elastic
interactions among the local distortions cause an effective long-range
interaction among the spin states.
The critical properties of the spin-crossover system turned out to
belong to the mean-field universality class.
It was also found that the finite-size analysis for various quantities 
is very similar to the long-range, weakly interacting model 
(the Husimi-Temperley model).

In spin-crossover systems, 
one may expect that the dynamics also corresponds to that of
the mean-field model.
In particular, various threshold phenomena have been pointed out in
the dynamics of spin-crossover type materials.
For example, a spinodal phenomenon without nucleation type clustering was
reported in numerical calculations \cite{Miyashita2008}.
The change is considered not to be a crossover process but a change with
a true critical singularity that can be described by 
mean-field dynamics \cite{Suzuki-Kubo1968}.
Moreover, threshold phenomena have been pointed out in the dependence 
of the photo-excitation process on the strength of the photo-irradiation, 
which can be modeled as a kind of spinodal phenomenon \cite{Miyashita2009}.
%These phenomena would be described by the mechanism of 
%spinodal decomposition in the mean-field dynamics \cite{Suzuki-Kubo1968}.
The metastable state does not relax to the stable state in the
mean-field approximation, which corresponds to the infinite system size in the
Husimi-Temperley model.
%%miya0426

In recent extensive studies on nano-size systems, finite-size effects 
turn out to play important roles.
%are among the most important properties.  
Therefore, it would be useful to study finite-size effects on the
spinodal phenomenon as a critical dynamical process.
The critical phenomena of the phase transition were studied
in a dynamical mean-field model. 
The relaxation time diverges when the parameter approaches 
the threshold value.
Binder studied the divergence of the relaxation time in 
spinodal phenomena in the mean-filed model by a Monte Carlo 
method~\cite{Binder1973}. 
Although 
the metastable state does not relax to the stable state in the
mean-field approximation, 
Paul, Heermann, and Binder studied 
the relaxation from the metastable side in finite-size systems, 
and obtained a finite-size scaling form of the relaxation time~\cite{Pauletal}.
The size dependence of the relaxation was also discussed recently \cite{Loscar2008}.

In the present paper, we study the asymptotic behavior of the relaxation 
time of the infinitely long-range interaction 
model (the Husimi-Temperley model) near the spinodal point 
including the metastable region, the spinodal point, 
and the unstable region.
We reexamine the scaling properties
near the spinodal point, which have been proposed
by Paul, Heermann, and Binder \cite{Pauletal}. 

As in the previous studies, %\cite{Binder1973,Pauletal},
we adopt the Glauber dynamics, and derive a master equation for the 
probability distribution of the total magnetization. 
We first derive a master equation as a function of the total
magnetization, which is possible in the Husimi-Temperley model because
of the long-range nature of the interactions.
Then, we derive a Fokker-Planck equation by using an expansion in terms
of the inverse system size, which is an example of the van Kampen
%%miya0426
Omega-expansion \cite{vanKampen,Kubo1973}.
We analyze the Fokker-Planck equation and derive a scaling relation and also 
asymptotic forms of the relaxation times. 
These properties are confirmed by direct 
numerical investigations of the original master equation, 
as well as corresponding Monte Carlo simulations of the stochastic decay process. 

We also extend the analysis to general threshold phenomena, such as 
switching from LS to HS states in spin-crossover materials by photo-irradiation.
We obtain the effects of the photo-irradiation on the master equation,
and show that the critical properties of these processes are the same as
those of the spinodal phenomena.

The rest of this paper is organized as follows.
In Section II, we review the spinodal phenomena and 
define the master equation for the long-range model. 
Then, we study the asymptotic size dependence of the relaxation 
time near the spinodal point in Section III. 
Numerical confirmation of the asymptotic forms is given in Section IV.
Next, we extend our study to threshold phenomena under external pumping in 
Section V. Finally, a summary and discussion are given in Section VI.

\section{Infinite long-range model and the spinodal point}

We investigate the relaxation phenomena near the spinodal point
in the infinitely long-range model.
This is a spin model in which each spin interacts equally 
with every other spin.  The Hamiltonian is
\beq
{\mathcal H}=-\frac{J}{2N}M^2-HM, \quad M=\sum_{i=1}^N \sigma_i ,
\label{eq:ham}
\eeq
where $H$ is the magnetic field and $\sigma_i=\pm 1$.
It is well known that the mean-field theory is exact for this model 
in the limit of $N\rightarrow\infty$.
This model shows a second-order phase transition at $T=T_{\rm C}=J$, $H=0$.
Below the critical temperature, a 
metastable state exists for weak magnetic fields.
When the magnetic field becomes strong, 
the metastable state becomes unstable at a certain point, known as 
the spinodal point.
In order to determine the spinodal point, we consider the extended 
free energy, i.e.,
the free energy for fixed total magnetization.
In the infinitely long-range model, this is given by
\begin{align}
f(m)&=-\frac{J}{2}m^2-Hm-\frac{1}{\beta N}\ln {}_N C _{(N+M)/2} \nonumber \\
&\sim -\frac{J}{2}m^2-Hm \nonumber \\
&+\frac{1}{\beta}\left(\frac{1+m}{2}\ln\frac{1+m}{2}
+\frac{1-m}{2}\ln\frac{1-m}{2}\right),
\label{eq:free}
\end{align}
in the limit $N\rightarrow\infty$
where we use Stirling's formula and $m$ is the magnetization per spin ($m=M/N$).
The spinodal point is given by the following conditions,
$$\frac{\d f}{\d m}=0 \quad {\rm and} \quad \frac{\d ^2 f}{\d m^2}=0 ,$$
which give
\beq
H_{\rm SP}=\mp J\rt{1-\frac{1}{\beta J}}
\pm\frac{1}{2\beta}\ln\frac{1+\rt{1-\frac{1}{\beta J}}}{1-\rt{1-\frac{1}{\beta J}}},
\eeq
at which the magnetization is given by
\beq
m_{\rm SP}=\pm\rt{1-\frac{1}{\beta J}}. 
\eeq 
In this paper, we consider the case where we increase the field from a 
negative value. Therefore, we consider the behavior of a locally stable 
state at negative magnetization around $m_{\rm SP}<0$.

%\section{The Dynamics near the spinodal point}
\label{sec:dynamics}

We study the dynamics via the standard master equation
\beq
\frac{\d P(S,t)}{\d t}=-\sum_{S'}W_{S\rightarrow S'}P(S)+\sum_{S'}W_{S'\rightarrow S}P(S'),
\label{eq:mastereq0}
\eeq
where $S$ and $S'$ denote states of the system 
and $W_{S\rightarrow S'}$ is a transition rate from  $S$ to $S'$. 
The probability of the state $S$ at time $t$ is denoted by $P(S,t)$.

Among the many possible dynamical models 
(choices of the transition rate), 
we adopt the Glauber dynamics in this work.
In the Glauber dynamics, the transition takes place as a flip of a local 
spin, and 
the transition rate $w_{ij}$ from a local spin state $i$ to a local spin state
$j$ is given by
\beq
w_{ij}=\frac{1}{\tau_0}{e^{-\beta E_j}\over e^{-\beta E_i}+e^{-\beta E_j}},
\label{eq:Glauber}
\eeq
where $E_i$ denotes the energy of the system in spin state $i$,
and $\tau_0$ is some characteristic time scale.
%%miya0426
In this paper, we scale the time by $\tau_0$ and set $\tau_0=1$ for simplicity.

With this transition rate, we construct a master equation for the 
mean-field model.
As the Hamiltonian depends only on the magnetization $M$,
the master equation is written in closed form for $M$.
Thus the master equation (\ref{eq:mastereq0}) 
can be expressed as a function of $M$. 
Let $P(M)$ be the probability that the system has the total magnetization $M$.
The master equation for $P(M)$ is given by
\begin{align}
&\frac{\d P(M,t)}{\d t}={1\over\tau_0}\times\nonumber \\
&\left\{-\frac{N+M}{2}\frac{\exp[-\beta(J(M-1)/N+H)]}{2\cosh[\beta(J(M-1)/N+H)]}P(M)\right.
 \nonumber\\
&-\frac{N-M}{2}\frac{\exp[\beta(J(M+1)/N+H)]}{2\cosh[\beta(J(M+1)/N+H)]}P(M)
 \nonumber \\
&+\frac{N-M+2}{2}\frac{\exp[\beta(J(M-1)/N+H)]}{2\cosh[\beta(J(M-1)/N+H)]}P(M-2)\nonumber \\
&\left. +\frac{N+M+2}{2}\frac{\exp[-\beta(J(M+1)/N+H)]}{2\cosh[\beta(J(M+1)/N+H)]}P(M+2)\right\},
\label{eq:master}
\end{align}
where $N$ is the number of spins and $M$ takes discrete values $-N, -N+2,\cdots N$ (see Appendix \ref{sec:der_ms}).
%%miya0426
%Here $\tau_0$ is the time scale of the dynamics. 
%In this paper, we scale the time by $\tau_0$ and set $\tau_0=1$ for simplicity.
For finite $N$ we can solve the simultaneous equations for $P(-N,t),
P(-N+2,t),\cdots, P(N,t)$, as well as perform a Monte Carlo simulation
of the model. 
%%miya0426
%Therefore, there is no statistical error bar, which will be given in the section IV.

\section{Asymptotic size-dependence of the relaxation time near 
the spinodal point}

Next, we study the scaling properties
of the relaxation time near the spinodal point.
For large $N$, we put $m=M/N$ and regard $m$ as a continuous variable.
Expanding the RHS of Eq.~(\ref{eq:master}) in a series in $\varepsilon =1/N$,
we obtain the following Fokker-Planck equation:
\beq
\frac{\d P(m)}{\d t}=\frac{\d}{\d m}g_1(m)P(m)+\varepsilon\frac{\d^2}{\d m^2}g_2(m)P(m)+O(\varepsilon^2),
\label{eq:FP}
\eeq
where
\beq
\begin{split}
g_1(m)&=m-\tanh[\beta(Jm+H)]+\varepsilon\frac{\beta Jm}{\cosh^2[\beta(Jm+H)]} \\
g_2(m)&=1-m\tanh\left[\beta(Jm+H)\right] .
\end{split}
\label{g1g2}
\eeq
The last term of $g_1(m)$ gives the correction to the spinodal point
due to the finite-size effect.
Hereafter, we ignore this term as it is very small.
Near the spinodal point, we expand $g_1(m)$ and $g_2(m)$ around the spinodal point.
We set $x=m-m_{\rm SP}$, and $y=\beta(H-H_{\rm SP})\equiv h-h_{\rm SP}$, and then we have 
\beq
\begin{split}
g_1(m)&\simeq -\frac{y}{\beta J}-\eta (\beta Jx^2+2xy) \\
g_2(m)&\simeq \frac{1}{\beta J},
\end{split}
\label{eq:g}
\eeq
where $\eta=|m_{\rm SP}|=\rt{1-1/\beta J}$.

Let us consider the time evolution of $x$, starting from $x=0$.
The distribution of $x$ evolves according to Eq.~(\ref{eq:FP})
with $g_1\sim -y/\beta J$ and $g_2\sim 1/\beta J$.
When $x$ approaches $x\sim y^{1/2}$,
the correction term $-\eta\beta Jx^2$ in $g_1$ becomes relevant,
but other correction terms in $g_1$, such as $-2\eta xy$, are still irrelevant.
Therefore, in order to determine the relaxation time,
we can use the approximation 
that $g_1\simeq -y/\beta J-\eta\beta Jx^2$ in the early stage of the phase 
change.  Then, the Fokker-Planck equation takes the form
\beq
\frac{\d}{\d t}P(x,t) = \left(
-\frac{\d}{\d x}\left(\frac{y}{\beta J}+\eta\beta Jx^2\right) 
+\frac{\varepsilon}{\beta J}\frac{\d^2}{\d x^2}
\right)P(x,t).
\label{eq:FP2}
\eeq
Now, we introduce the scaled parameters, 
\beq
\xi=x|y|^{-1/2},
\eeq
and 
\beq
\Lambda=N^{2/3}y.
\eeq
Then, for nonzero $y$, the Fokker-Planck equation is given by
\begin{align}
\frac{\d}{\d t}P(\xi,t) = \varepsilon^{1/3}&\left\{
-|\Lambda|^{1/2}\frac{\d}{\d \xi}\left(\pm\frac{1}{\beta J}+\eta\beta J\xi^2\right)\right.\nonumber \\
&\left. +\frac{1}{\beta J|\Lambda|}\frac{\d^2}{\d \xi^2}\right\}
P(\xi,t),
\label{eq:FP3}
\end{align}
where $\Lambda >0$ for the upper sign, and $\Lambda <0$ for the lower sign.
Because Eq.~(\ref{eq:FP3}) depends only on $\Lambda$
except for the factor $\varepsilon^{1/3}$,
the relaxation time is expected to be given in the form
\beq
\tau \sim N^{1/3}f(\Lambda) = N^{1/3}f(N^{2/3}(h-h_{\rm SP})).
\label{eq:scaling}
\eeq
This is the finite-size scaling for the relaxation time near the spinodal point.
This form has been pointed out by Paul, et al.\cite{Pauletal}.
Here, we reexamine the form of the scaling function by studying the asymptotic
forms of the relaxation time.

%%%
In the above argument leading to the scaling relation,
it is necessary to pay attention to the range of the parameters, 
in which the application of the above estimate can be verified.
We may regard the relaxation time as a time when the magnetization $x$ 
becomes $O(1)$. Then, $\xi$ becomes $O(|y|^{-1/2})$.
This implies that we may need to include an additional $y$-dependence in
the relaxation time. Thus, we cannot immediately conclude that the finite-size
scaling has the simple form of Eq.~(\ref{eq:scaling}).
Indeed, such consideration is essential for the relaxation after 
rapid quenching at zero field and has been studied as a scaling theory
of the relaxation at unstable point~\cite{Suzuki1976}.  
In fact, the relaxation time is proportional to $\ln N$ in the relaxation at the unstable point,
 which is unexpected from the form of the Fokker-Planck equation.
This problem is investigated in Appendix \ref{sec:cutoff}, where we 
show that the additional contribution does {\it not} need 
to be taken into account in the relaxation at the spinodal point,
 and the finite-size scaling is correctly given by Eq.~(\ref{eq:scaling}).
In the following, we investigate asymptotic forms of the relaxation time 
for the cases $y<0$, $y=0$, and $y>0$.

%090830
In order to consider more general cases later (Sec.~\ref{sec:pump}), 
we rewrite the Fokker-Planck equation with general coefficients $\alpha$, $\gamma$, and $\delta$ :
\begin{align}
\frac{\d}{\d t}P(\xi,t) = \varepsilon^{1/3}&\left\{
-|\Lambda|^{1/2}\frac{\d}{\d \xi}\left(\pm\alpha +\gamma\xi^2\right)\right.\nonumber \\
&\left. +\frac{\delta}{|\Lambda|}\frac{\d^2}{\d \xi^2}\right\}
P(\xi,t).
\label{eq:FP5}
\end{align}
The coefficients $\alpha$, $\gamma$, and $\delta$ are defined as follows:
\beq
\begin{split}
g_1&\simeq -\alpha y-\gamma x^2 \\
g_2&\simeq\delta .
\end{split}
\label{eq:g2}
\eeq
In the case of the relaxation from the mean-field spinodal point,
$\alpha=\delta=1/(\beta J)$, $\gamma=\eta\beta J$.

\subsection*{Metastable region: the $y<0$ case}

\begin{figure}[tbhp]
\begin{center}
\includegraphics[scale=0.5]{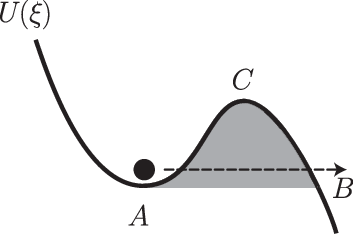}
\caption{
The scaled potential $U(\xi)$ in Eq.~(\protect\ref{eq:potential}). 
A: the metastable state. 
B: a point in the basin of attraction of the stable state. 
C: the unstable state. 
}
\label{fig:potential}
\end{center}
\end{figure} 

This case corresponds to the relaxation from the metastable state to the equilibrium state.
We can estimate the relaxation time by Kramers' method of escape over a potential barrier \cite{Kramers1940}.
We rewrite Eq.~(\ref{eq:FP5}) in the following form ($\Lambda <0$)
\beq
\frac{\d P}{\d t} = \varepsilon^{1/3}\frac{\d}{\d \xi}\left(
\frac{\delta}{|\Lambda|}e^{-|\Lambda|^{3/2}U(\xi)/\delta}
\frac{\d}{\d \xi}e^{|\Lambda|^{3/2}U(\xi)/\delta}P\right),
\label{eq:FP4}
\eeq
where 
\beq
U(\xi)=\alpha\xi -\gamma\xi^3/3
\label{eq:potential}
\eeq
is the scaled potential. 
We consider the escape of probability from the valley to the outside 
(A $\rightarrow$ B) as depicted in Fig.~\ref{fig:potential} \cite{Kramers1940}.
There, the probability current $\sigma$ is given by
\beq
\sigma =-\varepsilon^{1/3}\frac{\delta}{|\Lambda|}e^{-|\Lambda|^{3/2}U(\xi)/\delta}
\frac{\d}{\d \xi}e^{|\Lambda|^{3/2}U(\xi)/\delta}P.
\eeq
We consider the stationary current ($\sigma=$const.) 
and integrate it between the two points A and B.
\beq
\sigma=-\varepsilon^{1/3}\frac{\delta}{|\Lambda|}\frac{[e^{|\Lambda|^{3/2}U(\xi)/\delta}P]_A^B}
{\int_A^B e^{\beta J|\Lambda|^{3/2}U(\xi)/\delta}d\xi}.
\eeq
If the number of spins $N$ is very large, we can use the steepest-descent 
method, and we have 
\begin{align}
&\int_A^B e^{|\Lambda|^{3/2}U(\xi)/\delta}d\xi \nonumber \\
&\approx \int_{-\infty}^{\infty} 
e^{\frac{1}{\delta}|\Lambda|^{3/2}U(C)+\frac{1}{2\delta}\beta J|\Lambda|^{3/2}U''(C)(\xi-C)^2}d\xi \nonumber \\
&=\rt{\frac{-2\pi\delta}{|\Lambda|^{3/2}U''(C)}}e^{|\Lambda|^{3/2}U(C)/\delta}.
\end{align}
This approximation is valid for a sufficiently large $|\Lambda|$.
We consider the early stage of the relaxation and 
assume the relaxation has not occurred yet, namely $P(B)\approx 0$. 
Then, we obtain the estimate:
\beq
\sigma=P(A)\varepsilon^{1/3}
\rt{\frac{-\delta U''(C)}{2\pi|\Lambda|^{1/2}}}e^{-|\Lambda|^{3/2}(U(C)-U(A))/\delta}.
\label{eq:current}
\eeq
The probability distribution near the point A is approximately given by
\beq
P(\xi)\simeq n_{\rm A}\frac{\exp\left[-\frac{|\Lambda|^{3/2}}{\delta}
\left\{U(A)+\frac{1}{2}U''(A)(\xi -A)^2\right\}\right]}
{Z},
\label{eq:prob}
\eeq
where $Z$ is a partition function,
\begin{align}
Z&=\int_{-\infty}^{\infty}d\xi \exp\left[-\frac{|\Lambda|^{3/2}}{\delta}\left\{U(A)
+\frac{1}{2}U''(A)(\xi -A)^2\right\}\right] \nonumber \\
&=e^{-|\Lambda|^{3/2}U(A)/\delta}\rt{\frac{2\pi\delta}{|\Lambda|^{3/2}U''(A)}}.
\end{align}
Namely, the probability distribution near the point A is given by the 
equilibrium distribution of the approximate harmonic potential.
The variable $n_{\rm A}$ represents the total probability near the point A.
This quantity evolves as
\beq
\frac{d}{dt}n_{\rm A}=-\sigma =-\frac{1}{\tau}n_{\rm A} ,
\label{eq:nA}
\eeq
where $\tau$ is the relaxation time that we want to know.
As the probability at the point A is given by
\begin{align}
P(A)&\simeq n_{\rm A}\frac{\exp\left[-\frac{|\Lambda|^{3/2}}{\delta}U(A)\right]}{Z}\nonumber \\
&=n_{\rm A}\rt{\frac{|\Lambda|^{3/2}U''(A)}{2\pi\delta}},
\label{eq:prob_A}
\end{align}
(see Eq.(\ref{eq:prob})), combining Eqs.(\ref{eq:current}), (\ref{eq:nA}), and (\ref{eq:prob_A}), we obtain
\begin{align}
\tau \sim N^{1/3}2\pi |U''(A)U''(C)|^{-1/2}|\Lambda|^{-1/2}\nonumber \\
\times\exp\left[\frac{|\Lambda|^{3/2}}{\delta}(U(C)-U(A))\right]
\;.
\label{eq:meta_Kramers}
\end{align}
This is the result of the well-known Kramers' formula for the escape rate, 
and it agrees with the finite-size scaling equation (\ref{eq:scaling}).

The potential $U(\xi)$ is $U(\xi)=\alpha\xi-\gamma\xi^3/3$,
and the two points A and C are given by the condition $dU/d\xi=0$.
Therefore,
\beq
\begin{split}
A=-C&=-\frac{\alpha}{\gamma}, \\
U(C)-U(A)&=\frac{4\alpha}{3}\rt{\frac{\alpha}{\gamma}},\\
U''(C)&=-2\rt{\alpha\gamma}.
\end{split}
\eeq
Hence, the relaxation time for sufficiently large $|\Lambda|$ is
\beq
\tau \sim N^{1/3}\frac{\pi}{2\rt{\alpha\gamma|\Lambda|}}
\exp\left\{\frac{4\alpha}{3\delta}\rt{\frac{\alpha}{\gamma}}|\Lambda|^{3/2}\right\}.
\label{eq:tau_meta}
\eeq

In the case of Eq.~(\ref{eq:g}), $\alpha=\delta=1/(\beta J)$, $\gamma=\eta\beta J$, and we have
\beq
\tau\sim N^{1/3}\frac{\pi}{2\rt{\eta|\Lambda|}}
\exp\left\{\frac{4}{3\beta J\eta^{1/2}}|\Lambda|^{3/2}\right\}.
\eeq
Here, it should be noted that from Eq.~(\ref{eq:free})
the following relation holds: 
$$\frac{4}{3\beta J\eta^{1/2}}\Lambda^{3/2} =\beta\Delta F \equiv \beta (F(C)-F(A)).$$
Therefore, the relaxation time obtained above is roughly 
$\tau \sim e^{\beta\Delta F}$, which is the well-known Arrhenius formula.
Another derivation of Eq.~(\ref{eq:tau_meta}) uses the WKB approximation 
as discussed by Tomita, et al.\ \cite{Tomita1976}. 
We can derive the same result %(not shown).
(see Appendix \ref{sec:WKB}). 

In the infinite long-range model, microscopic fluctuations do not 
grow to become macroscopic because the long-range interaction suppresses 
clustering and nucleation. It should be noted that, in the 
limit of $N\rightarrow\infty$, the system remains at the metastable 
or marginally stable point, and the relaxation time from that 
point becomes infinite. 
In contrast, in systems with short-range interactions, the nucleation 
process causes the system to relax to the equilibrium state in a finite 
relaxation time. Thus, the divergence of the relaxation time 
does not take place, and so far the divergence of the 
relaxation time has not been considered seriously. However, as has 
been pointed out \cite{Miyashita2008}, effective long-range interactions 
appear in systems in which elastic deformation mediates interactions 
among the spins. In such systems, the long-range interaction model is 
effectively realized, and we expect that the finite-size scaling 
discussed here would be relevant.

\subsection*{At the spinodal point: the $y=0$ case} %%%%%%%%%%%%%%%%%%%%%%%%%%%%%%%%%
Next, we consider the relaxation just at the spinodal point, $y=0$.
Substituting $y=0$ in the Fokker-Planck equation (\ref{eq:FP2}), we obtain
\beq
\frac{\d}{\d t}P(x,t)=\gamma\frac{\d}{\d x}\left(x^2P(x,t)\right)
+\varepsilon\delta\frac{\d^2}{\d x^2}P(x,t).
\eeq
Putting $x=\varepsilon^{1/3}z$,
\beq
\frac{\d}{\d t}P(z,t)=\varepsilon^{1/3}\left\{
-\gamma\frac{\d}{\d z}z^2+\delta\frac{\d^2}{\d z^2}\right\} P(z,t). 
\eeq
By using the scaled variable $s=t\varepsilon^{1/3}$, we can eliminate the $\varepsilon$-dependence. 
Thus, as pointed out by Kubo et al.\ \cite{Kubo1973},
the relaxation time behaves as 
\beq
\tau \propto \varepsilon^{-1/3}=N^{1/3}.
\eeq
The relaxation time diverges in the limit of $N\rightarrow\infty$ just at 
the spinodal point. In the limit of $N\rightarrow\infty$, the system 
remains at the unstable point, and the relaxation time becomes infinite. 
This divergence is again due to the long-range interaction.

\subsection*{Unstable region: the $y>0$ case} %%%%%%%%%%%%%%%%%%%%%%%%%%%%%%%%%%%%%

Finally, we consider the case $y>0$.
In this case, even if $N=\infty$ ($\varepsilon=0$), the relaxation 
takes place. Namely, the relaxation time saturates at a finite value 
at large $N$. Therefore, we consider only the limit $N\rightarrow\infty$. 
The Fokker-Planck equation (\ref{eq:FP3}) then becomes
\beq
\frac{\d P}{\d t}=-y^{1/2}\frac{\d}{\d\xi}\left(\alpha+\gamma\xi^2\right)P.
\eeq
Therefore, the relaxation time is expected to scale as
\beq
\tau \sim y^{-1/2}\sim (h-h_{\rm SP})^{-1/2}.
\eeq
In the limit of $N\rightarrow\infty$, there is no diffusion term,
so we can derive the time evolution of the scaled magnetization $\xi(t)$ directly. Namely, putting $P(\xi ,t)=\delta(\xi-\xi(t))$, we obtain
\beq
\dot{\xi}(t)=y^{1/2}\left(\alpha+\gamma \xi(t)^2\right).
\label{eq:unstable}
\eeq
The solution of Eq.(\ref{eq:unstable}) is given by
\beq
\xi(t)=\rt{\frac{\alpha}{\gamma}}\tan \left(\rt{\alpha\gamma y}t\right)
\eeq
for $\rt{\alpha\gamma y}t<\pi/2$.
At a time $\rt{\alpha\gamma y}t=\pi/2$, the above expression indicates that $\xi(t)$ would diverge.
However, the higher order terms in the original Fokker-Planck equation (\ref{eq:FP}) prevent this divergence of the magnetization.
Thus, the relaxation time is estimated as
\beq
\tau \sim \frac{\pi}{2}(\alpha\gamma y)^{-1/2}.
\eeq
In the scaling form,
\beq
\tau \sim N^{1/3}\frac{\pi}{2}(\alpha\gamma\Lambda)^{-1/2}.
\label{Eq40}
\eeq
This result is consistent with that obtained by Binder \cite{Binder1973}, who
showed that for $h>h_{\rm SP}$, the relaxation time behaves as 
$\tau \sim (h-h_{\rm SP})^{-1/2}$.

\section{Numerical results} %====================================

We have seen that the relaxation time $\tau$ obeys the finite-size scaling form 
(\ref{eq:scaling}):
$$\tau(y,N)=N^{1/3}f(N^{2/3}y),$$
and we have derived the asymptotic forms of the relaxation time, i.e., in the case of Eq.~(\ref{eq:g}),
\beq
f(\Lambda)\sim\left\{
\begin{split}
 &\pi (\eta|\Lambda|)^{-1/2}
\exp\left\{\frac{4}{3\beta J\eta^{1/2}}|\Lambda|^{3/2}\right\}
\ {\rm for} \ -\Lambda \gg 1 \\
&\frac{\pi}{2}(\eta\Lambda)^{-1/2} \quad {\rm for} \quad \Lambda \gg 1
\end{split}
\right.
\label{eq:asymptotic}
\eeq

%\begin{figure}[tbhp]
%\begin{center}
%\includegraphics[scale=0.4, angle=90]{sp-scaling2.eps}
%\caption{Scaling plot of the relaxation time in the coordinate 
%($\Lambda=N^{2/3}y$, $\ln(N^{-1/3}\tau)$) for the parameters $\beta=1$ 
%and $J=2$. All the data collapse well onto a scaling function which should 
%be $\ln f(\Lambda)$.}
%\label{fig:scaling}
%\end{center}
%\end{figure} 

\begin{figure}[tbhp]
\begin{center}
\includegraphics[scale=0.5,angle=0]{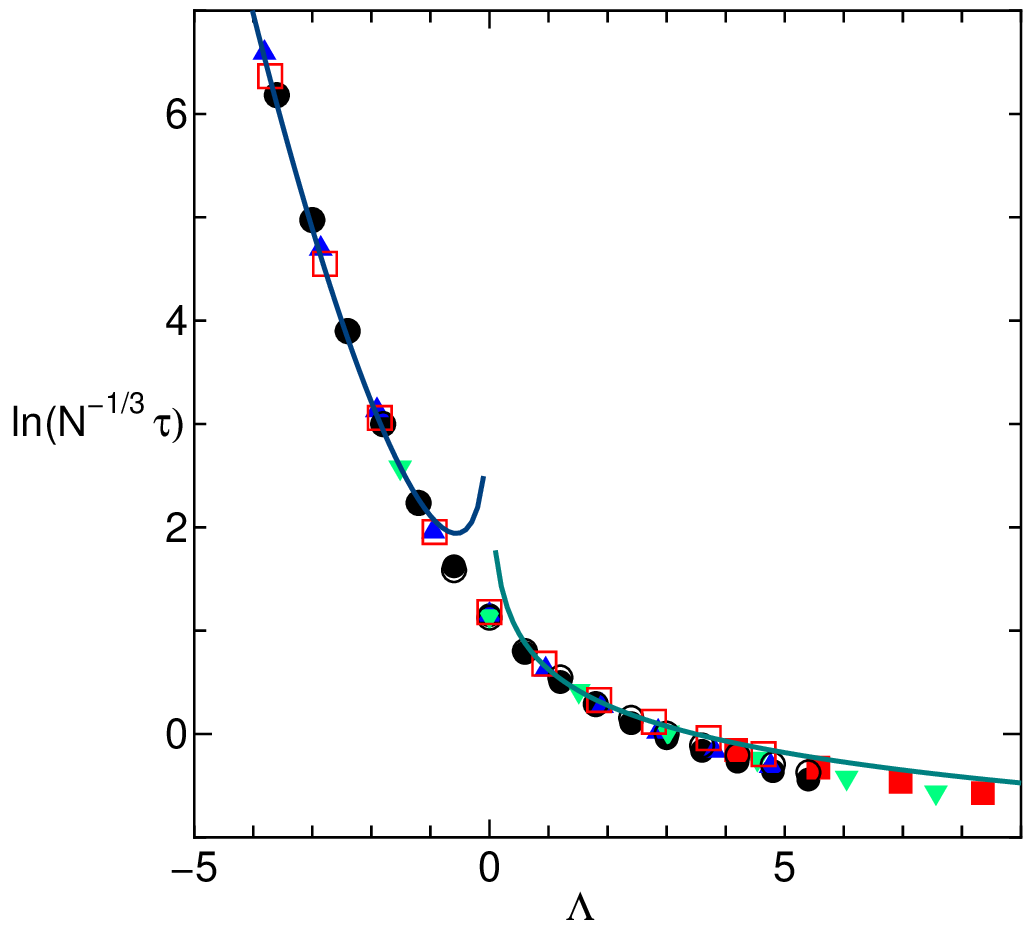}
\caption{(Color online) 
Comparison of the asymptotic form of the relaxation time with 
numerical results.
Data are plotted in the coordinates ($\Lambda$,  $\ln(N^{-1/3}\tau)$).
%ME and MC represent the solutions of the master equation and
%the Monte Carlo results, respectively.
The solid lines denote the asymptotic form (\ref{eq:asymptotic}).
%The symbols denote as follows: 
The symbols are interpreted as follows: 
Closed circles, upward triangles, downward triangles and closed squares denote 
the data obtained by the master equation 
for $N=1000$, 2000, 4000 and 10000, respectively.
Open circles and squares denote the data obtained by the Monte carlo method for
$N=1000$ and 10000, respectively.
}
\label{fig:scaleplot}
\end{center}
\end{figure}

In this section, we check these results by numerical studies.
We solved the original master equation (\ref{eq:master}) numerically, 
and also performed kinetic MC simulations (see Appendix D).  
%%%0411, for $N=1000,2000$ and $4000$.
The parameters are set as $\beta=1$ and $J=2$.
The relaxation time is defined as the time at which 
the magnetization of a sample reaches a certain value $m_0$.
Here we adopt $m_0=0$. 
In the Monte Carlo simulations, the relaxation time is measured 
directly in each simulation. 
On the other hand, in the master equation we
have to define it from the change of the probability distribution $P(M,t)$.
Namely, we obtain the average of the relaxation time with the
formula,
\beq
\tau = -\int_0^{\infty}dt \sum_{M<0}\dot{P}(M,t)t,
\eeq
and its standard deviation as 
\beq
\sigma_\tau = -\int_0^{\infty}dt \sum_{M<0}\dot{P}(M,t)t^2-\tau^2. 
\eeq
We plot data for various parameters in a scaling plot in 
Fig.~\ref{fig:scaleplot}, namely, in the coordinates
($\Lambda=N^{2/3}y$, $\ln(N^{-1/3}\tau)$). 
We confirmed that both methods give the same results, as they should.  
In the Monte Carlo simulations, 
each data point is an average over 1000 samples, 
and the error bars are smaller than the symbol size in the figure.
All the data collapse well onto a scaling function, 
which indicates that the finite-size scaling works well.
For large $\Lambda$, data points for different $N$ deviate 
slightly from the scaling function. 
This fact is explained as follows:
The condition for the finite-size scaling to hold is that the system 
size $N$ is sufficiently large and the magnetic field is sufficiently 
close to the spinodal point. This implies $N\gg 1$ and $|y|\ll 1$.
However, an even stronger condition is required for the finite-size scaling.
As we assumed $x=m-m_{\rm SP}\ll 1$ to derive the Fokker Planck equation and
$x$ was rescaled as $x=\xi |y|^{1/2}$, not only $|y|\ll 1$ but also 
$|y|^{1/2}\ll 1$ was necessary.
Therefore, 
$$|\Lambda|^{1/2}\ll N^{1/3}$$
is required for the finite-size scaling.

\begin{figure}[tbhp]
\begin{center}
\includegraphics[scale=0.4,angle=0]{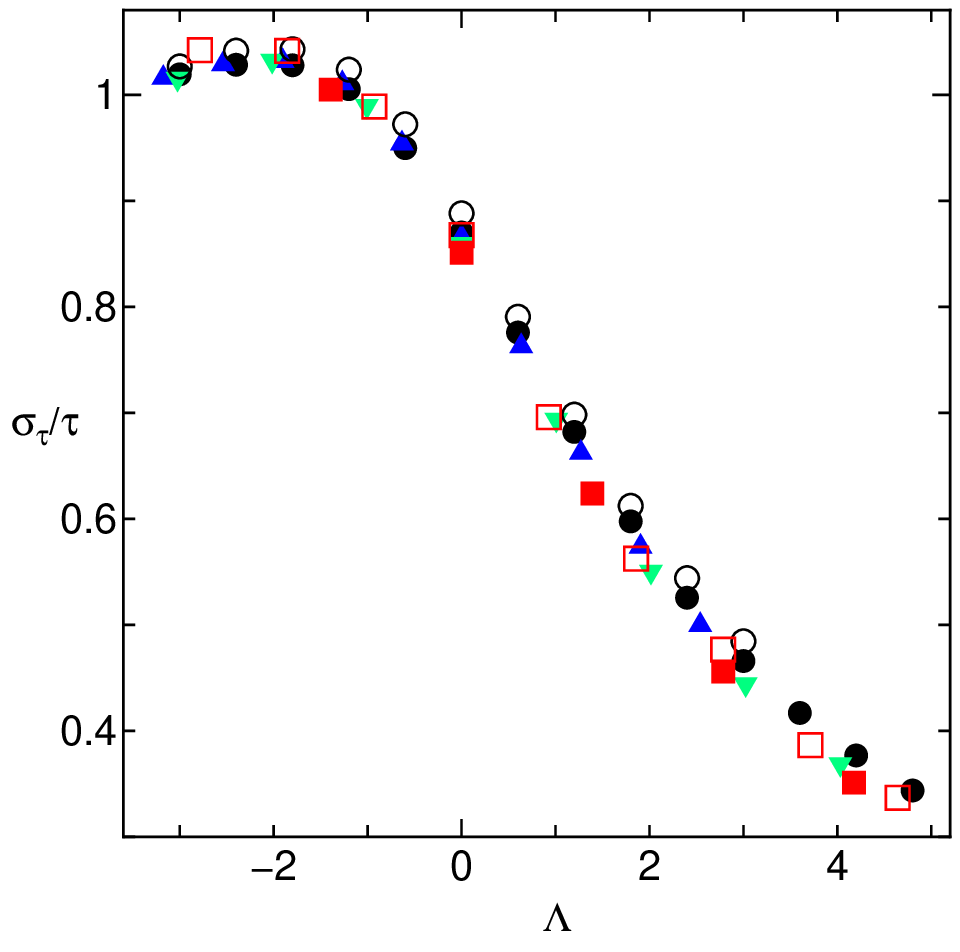}
\caption{(Color online)
Ratio of the standard deviation of the lifetime, 
$\sigma_\tau$, to the lifetime, $\tau$, as obtained from 
the master equation and MC simulations. 
The data are shown as a scaling plot vs the scaling variable $\Lambda$.
%The symbols denote the same as in Fig.~\ref{fig:scaleplot}.}
The symbols have the same interpretations as in Fig.~\ref{fig:scaleplot}.}
\label{fig-per}
\end{center}
\end{figure}

Now, we compare the numerical results of the master equation with the 
asymptotic form for the relaxation time $\tau$, Eq.~(\ref{eq:asymptotic}).
In Fig.~\ref{fig:scaleplot}, we compare numerical results and the 
asymptotic forms, Eq.~(\ref{eq:asymptotic}).
Here we find that the asymptotic formulae describe the 
scaling form well in the large-$\Lambda$ region, 
where the data points approach the asymptotic forms 
when the size increases.  
Here, we find that the scaling property (\ref{eq:scaling}) holds, and 
the asymptotic forms also hold asymptotically.

%%%%%%%%%%%%%%%%%%%%%%%%%%%%%%%%%%%%%%%%%%%%%%%%%%%%%%%%%%%%%%%%1018

As the scaling variable $\Lambda$ goes from large positive to large 
negative values, 
the lifetime distribution goes from a narrow peak in the unstable region 
to an exponential distribution deep in the metastable region. 
This is illustrated in Fig.~\ref{fig-per} by the ratio of the standard 
deviation 
of the lifetime, $\sigma_\tau$, to the lifetime, $\tau$, as obtained 
from the Master equation and also by MC simulations. 
These results also show good scaling collapse, 
%XXX Per 11/08/09:
although the convergence to the 
asymptotic limit is somewhat slower for $\sigma_\tau$ than for $\tau$, 
as one would expect for a quantity involving a second moment of the lifetime 
distribution.

%%%%%%%%%%%%%%%%%%%%%%%%%%%%%%%%%%%%%%%%%%%%%%%%%%%%%%%%%%%%%%%%%1018

\section {Threshold phenomena under external pumping}
\label{sec:pump}

In this section we study switching in spin-crossover (SC) materials under photo-irradiation, 
which is representative of a wide range of switching processes under external driving forces.
The pumping effect is described by the following process in the master equation.
The pumping operation causes a down spin (LS) to flip to the up state (HS) 
with a transition rate $a$ per unit time.
Thus, the total transition rate (\ref{eq:Glauber}) from down to up 
is now given by
\beq
w(-\rightarrow +)={1\over \tau_0}{e^{-\beta E_{+}}\over 
e^{-\beta E_{+}}+e^{-\beta E_{-}}}+a,
\eeq 
where $E_{\pm}$ denote the energies with the spin $+$ and $-$, respectively.
On the other hand, the transition rate from up to down remains unchanged and 
is given by
\beq
w(+\rightarrow -)={1\over \tau_0}{e^{-\beta E_{-}}\over 
e^{-\beta E_{+}}+e^{-\beta E_{-}}}.
\eeq 
These transition rates do not satisfy detailed balance, but
they produce a stationary state. Thus, as far as we consider the system at a
given temperature, we have a well-defined relaxation process to the 
stationary state. Here, we study the parameter dependence of the relaxation time.
To express the pumping, we add the term
\beq
-a{N-M\over2}P(M)+a{N-(M-2)\over2}P(M-2)
\eeq
in the master equation. 
In the large $N$ limit, this modification gives the additional terms
in the Fokker-Planck equation
\beq
-a{\partial\over\partial m}\left\{(1-m)P(m,t)\right\}
+a\varepsilon{\partial^2\over\partial m^2}\left\{(1-m)P(m,t)\right\}.
\eeq
Thus $g_1(m)$ and $g_2(m)$ in Eq.~(\ref{g1g2}) are now replaced by
\beq
\begin{split}
g_1(m)&=m-\tanh[\beta(Jm+H)]-a(1-m) \\
g_2(m)&=1-m\tanh\left[\beta(Jm+H)\right] +a(1-m).
\end{split}
\eeq
The point corresponding to the spinodal point is
given by
\beq
g_1(m)=0 \quad {\rm and} \quad {\partial\over\partial m}g_1(m)=0,
\eeq
which are given by
\beq
\tanh z=\frac{1+a_c}{\beta J}z-\frac{H}{J}(1+a_c)-a_c,
\label{cond1}
\eeq 
and
\beq
\cosh^2 z={\beta J\over 1+a_c},
\label{cond2}
\eeq 
where $z=\beta (J m_{\rm sp}+H)$.
%%%miya1018
For the SC material, $H$ consists of the following two ingredients:
the crystal field which gives the energy difference between HS and LS states,
and a time-dependent field which represents the different degeneracies 
of the HS and LS states~\cite{miyaPTP}.   

Expanding the variable near the spinodal point: 
$x=m-m_{\rm sp}$ and $y=a-a_c$, we have
\beq
g_1\simeq -\alpha y-\gamma x^2,
\eeq
where
\beq
\alpha={1\over 1+a_c}\left(1-\sqrt{1-{1+a_c\over\beta J}}\right),
\eeq
and 
\beq
\gamma=\beta J(1+a_c)\sqrt{1-{1+a_c\over\beta J}},
\eeq
and
\beq
g_2\simeq \frac{1}{\beta J}+\frac{2a_c}{1+a_c}\left(1+\rt{1-\frac{1+a_c}{\beta J}}\right)
\equiv\delta .
\eeq
As $g_1$ and $g_2$ have the same forms as Eq.~(\ref{eq:g2}), 
the same analysis as in the previous section remains valid,
and the finite-size scaling is again given in the form
\beq
\tau=N^{1/3}f(N^{2/3}(a-a_c)).
\eeq
The asymptotic form is also given by Eqs.~(\ref{eq:tau_meta}) and (\ref{Eq40}). 
%\beq\tau=\left\{\begin{array}{ccl} 
%N^{1/3}{\pi\over 2\sqrt{\alpha \gamma\Lambda}} \quad (\Lambda \gg 1)\\
%N^{1/3}{\pi\over 2\sqrt{\alpha \gamma\Lambda}}\exp\left(-{4\alpha\over3\delta}
%\sqrt{\alpha\over\gamma} |\Lambda|^{3/2}\right)\quad (\Lambda \ll -1)\end{array}\right.
%\eeq

\section{Summary and discussion}

Mean-field type 
critical behavior takes place in systems with effective long-range 
interactions, as has been pointed out for spin-crossover type 
materials \cite{Miyashita2008}. 
We expect that the dynamical critical properties such as the spinodal
phenomena are realized in those systems.
In models with short-range interactions,
there exists a mode of relaxation from the metastable state 
through nucleation of localized clusters. Thus, the relaxation time around the 
spinodal point changes smoothly, and the critical properties at the 
spinodal point in the mean-field theory are smeared out.
However, in long-range interaction models, the relaxation time
diverges as described by the mean-field theory.
It is, therefore, necessary to study the finite-size scaling 
properties of the critical behavior. Thus, we here studied 
the size dependence of the relaxation time near the spinodal point in the 
Husimi-Temperley model. Starting from the master equation for the 
probability density of the total magnetization 
under the Glauber dynamics, the Fokker-Planck 
equation for large $N$ was obtained as in previous work~\cite{Pauletal}. 
Using this Fokker-Planck equation, 
we investigated the relaxation processes near the spinodal point.
As a result, we obtained asymptotic forms and
a finite-size scaling function for 
the relaxation time, which cover both sides of the spinodal point, 
i.e., the metastable side and the unstable side.

We further extended the analysis to systems that are pumped by an external
field. The critical properties obtained in the present work should apply 
widely to 
threshold phenomena in long-range interacting models, such as the threshold 
phenomena found in the excitation process by photo-irradiation from the 
low-temperature phase to a photo-excited high-temperature phase 
in spin-crossover materials \cite{Miyashita2009}.

In the ongoing development of nano-size fabrication, switching processes in
nano-size systems are an important issue, and the size-dependence of relaxation
times of the switching processes must be precisely understood.
We hope the scaling properties presented here will help to analyze 
such processes in experimental systems.

\section*{Acknowledgments}
%%miya0426
The authors thank Professor William Klein for stimulating 
discussions on mean-field dynamics.
The authors would also like to thank Dr. Shu Tanaka for his helpful comments and discussions.
The present work was supported by Grant-in-Aid for Scientific Research
on Priority Areas, and also and the Next Generation Super Computer
Project, Nanoscience Program from MEXT of Japan.
The numerical calculations were supported by the supercomputer center of
ISSP of Tokyo University.
Work at Florida State University was supported by U.S.\ NSF Grant 
No.\ DMR-0802288. 

\appendix
\section{The derivation of the master equation}
\label{sec:der_ms}

In this Appendix, we derive the master equation (\ref{eq:master}) for the 
Hamiltonian (\ref{eq:ham}) and the transition rate (\ref{eq:Glauber}).
The probability of the state $\{\sigma_1,\sigma_2,\cdots ,\sigma_N\}$ at a time $t$,
which is denoted by $P(\sigma_1,\cdots ,\sigma_N ;t)$, evolves according to
\begin{align}
\frac{\d}{\d t}&P(\sigma_1,\cdots ,\sigma_N ;t)=
-\sum_{i=1}^N \omega_M(\sigma_i\rightarrow -\sigma_i)P(\sigma_1,\cdots ,\sigma_N ;t)\nonumber \\
&+\sum_{i=1}^N\omega_{M-2\sigma_i}(-\sigma_i\rightarrow\sigma_i)P(\cdots ,-\sigma_i,\cdots ;t)
\label{eq:motion}
\end{align}
where
\beq
\omega_M(\sigma_i\rightarrow -\sigma_i)=
\frac{1}{\tau_0}\frac{\exp\left(-\beta\sigma_i(J(M-\sigma_i)/N+H)\right)}
{2\cosh\left(\beta(J(M-\sigma_i)/N+H)\right)},
\eeq
and $M$ is the total magnetization, i.e., $M=\sum_i\sigma_i$.
We consider the time evolution of the probability of $M$:
$$P(M,t)\equiv \sum_{\sigma_1,\sigma_2,\cdots,\sigma_N=\pm 1}
\delta\left(\sum_{i=1}^N\sigma_i,M\right)P(\sigma_1,\cdots,\sigma_N;t),$$
where $\delta(a,b)$ denotes the Kronecker delta.
After some calculation from Eq.~(\ref{eq:motion}), 
the equation of motion for $P(M,t)$ is obtained in the form
\begin{align}
\frac{\d}{\d t}P(M,t)=
&-\frac{N+M}{2}\omega_M(+1\rightarrow -1)P(M,t)\nonumber \\
&-\frac{N-M}{2}\omega_M(-1\rightarrow +1)P(M,t)\nonumber \\
&+\frac{N-(M-2)}{2}\omega_{M-2}(-1\rightarrow +1)P(M-2,t)\nonumber \\
&+\frac{N+(M+2)}{2}\omega_{M+2}(+1\rightarrow -1)P(M+2,t).
\end{align}
The meaning of this equation is clear.
The first and second terms correspond to the transition from the magnetization $M$ 
to $M-2$ and $M+2$ respectively.
The third and fourth terms represent the transitions from $M-2$ to $M$ and from $M+2$ to $M$, respectively.
This equation gives Eq.~(\ref{eq:master}).

\section{Cut-off independence of the Fokker-Planck equation (\ref{eq:FP5})}
\label{sec:cutoff}

In Sec.~\ref{sec:dynamics}, we remarked on the possibility of an 
additional $y$-dependence in the relaxation time.
Namely, if we regard the relaxation time as the time when the magnetization $x$ becomes $O(1)$,
this corresponds to the time when $\xi$ becomes $O(|y|^{-1/2})$, and
this implies that we cannot conclude the finite-size scaling of the 
relaxation time, Eq.~(\ref{eq:scaling}),
from the form of the Fokker-Planck equation (\ref{eq:FP5}).
In other words, although the Fokker-Planck equation (\ref{eq:FP5}) seems to depend only on $\Lambda$,
we must restrict the range of the variable $\xi <O(y^{-1/2})$ and
this {\it cut-off} of $\xi$ can induce an additional $y$-dependence in the relaxation time.
We note that the relaxation time indeed depends on the cut-off in other situations.
One example is the relaxation from the mean-field unstable fixed point.
In this case, the Fokker-Planck equation is given by
\beq
\frac{\d}{\d t}P(x,t) = -\frac{\d}{\d x}xP(x,t)
+\varepsilon\frac{\d^2}{\d x^2}P(x,t),
\eeq
(we set some coefficients equal to unity).
We can transform this equation to the scaling form similarly.
If we set $x=\varepsilon^{1/2}\xi$,
\beq
\frac{\d}{\d t}P(\xi ,t) = -\frac{\d}{\d\xi}\xi P(\xi ,t)
+\frac{\d^2}{\d \xi^2}P(\xi ,t).
\label{eq:unstable_version}
\eeq
This equation is apparently independent of $\varepsilon$.
Is the relaxation time independent of the system size $N=1/\varepsilon$?
The answer is No.
It is known that the relaxation time in this case is $\tau\sim\ln N$ \cite{Suzuki1976}.
We show that this $N$-dependence stems from the finite cut-off.
We can solve Eq.~(\ref{eq:unstable_version}) for the initial 
condition $P(\xi ,0)=\delta(\xi)$,
\beq
P(\xi ,t)=\frac{1}{\rt{2\pi}\sigma(t)}\exp\left[-\frac{\xi^2}{2\sigma(t)^2}\right]
\eeq
where $\sigma(t)$ is given by
\beq
\sigma(t)=\rt{e^{2t}-1}\sim e^t
\eeq
It takes infinite time for $\xi$ to reach infinity.
As $x=\varepsilon^{1/2}\xi$, $\overline{x^2}(t)\sim \varepsilon\sigma(t)^2=\varepsilon e^{2t}$.
We consider the relaxation time as the time when $\overline{x^2}(t)$
reaches 1, i.e. $\overline{x^2}(\tau)\sim 1$,
the relaxation time is proportional to $\ln N$,
\beq
\tau \sim \ln N.
\eeq
In this way, we found out that the cut-off dependence could actually
affect the relaxation time,
but this cut-off played no role in the case of the relaxation near the spinodal point.

Hence, here we show that the relaxation time does not 
depend on the cut-off of $\xi$ if this cut-off is very large.

If we denote the average of $\xi^n$ over $P(\xi) $by $\overline{\xi^n}$,
the time evolution of $\overline{\xi}$ is given by
\begin{align}
\dot{\overline{\xi}}(t)&=\varepsilon^{1/3}|\Lambda|^{1/2}
\left(\pm\alpha+\gamma\overline{\xi^2}(t)\right) \nonumber \\
&\geq \varepsilon^{1/3}|\Lambda|^{1/2}
\left(\pm\alpha+\gamma\overline{\xi}(t)^2\right).
\label{eq:cutoff}
\end{align}
If $\xi(t_0)$ is larger than $\rt{\alpha/\gamma}\equiv\nu$, it can be shown from 
Eq.~(\ref{eq:cutoff}) that
\beq
\overline{\xi}(t)\geq \frac{1+A(t)}{1-A(t)}\nu,
\eeq
where $A(t)$ is given by
\beq
A(t)=\frac{\overline{\xi}(t_0)-\nu}
{\overline{\xi}(t_0)+\nu}\exp\left(2\varepsilon^{1/3}\Lambda^{1/2}\nu\gamma(t-t_0)\right).
\eeq
The average of the scaled magnetization $\overline{\xi}(t)$ reaches infinity 
when $A(t)=1$, namely
\beq
t-t_0=\frac{1}{2\varepsilon^{1/3}|\Lambda|^{1/2}\nu\gamma}
\ln\left(\frac{\overline{\xi}(t_0)+\nu}{\overline{\xi}(t_0)-\nu}\right).
\eeq
Because $t_0$ is finite, it takes only a finite time for $\overline{\xi}(t)$ 
to reach infinity.
Therefore, there is no cut-off dependence on the relaxation time in the Fokker-Planck equation (\ref{eq:FP5}).

\section{Derivation of Eq.~(\ref{eq:tau_meta}) by the WKB approximation}
\label{sec:WKB}

\begin{figure}[tbhp]
\begin{center}
\includegraphics[scale=0.5]{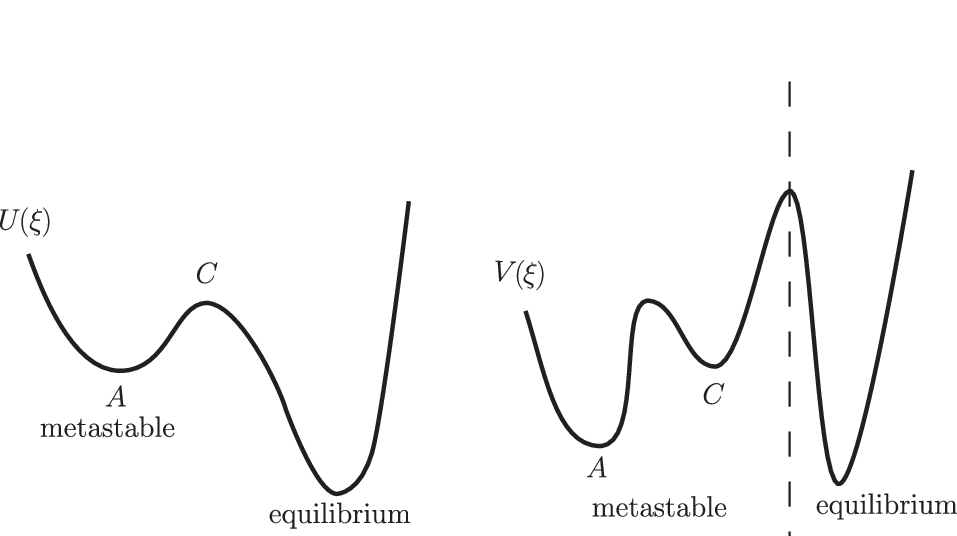}
\caption{Rough sketches of the scaled potential $U(\xi)$ and the corresponding
Schr\"odinger potential $V(\xi)$.}
\label{fig:potential2}
\end{center}
\end{figure} 
In the body of this paper, we 
estimated the relaxation time for $\Lambda <0$ and $|\Lambda|\gg 1$ 
according to Kramers' argument.
Here we give another derivation by using the WKB approximation.
The following derivation is essentially the same as that of 
Tomita, et al.\ \cite{Tomita1976}.
The Fokker-Planck equation (\ref{eq:FP4}) can be transformed to
the ``Schr\"odinger equation'' 
\beq
\frac{\d Q}{\d t}=\varepsilon^{1/3}\frac{\delta}{|\Lambda|}\frac{\d^2 Q}{\d\xi^2}
-\varepsilon^{1/3}V(\xi)Q(\xi)\equiv\varepsilon^{1/3}{\cal H}Q(\xi)
\label{eq:sch}
\eeq
by substituting
$$P=\exp\left(-\frac{1}{2\delta}|\Lambda|^{3/2}U(\xi)\right)Q(\xi).$$
The scaled potential $U(\xi)$ is given by Eq.~(\ref{eq:potential}), and
the Schr\"odinger potential $V(\xi)$ is 
\beq
V(\xi)=\frac{1}{4\delta}|\Lambda|^2U'(\xi)^2-\frac{1}{2}|\Lambda|^{1/2}U''(\xi).
\eeq
If the eigenvalues of ${\cal H}$ are $\lambda_i$ ($i=0,1,2,\cdots$) and the
eigenfunctions are $\phi_i$, we can expand $Q(\xi,t)$ as
\beq
Q(\xi,t)=\sum_ic_i\phi_i(\xi)e^{-\varepsilon^{1/3}\lambda_it}.
\eeq
The lowest eigenvalue is $\lambda_0=0$, and the corresponding eigenfunction is
\beq
\phi_0(\xi)= \frac{1}{Z^{1/2}}e^{-|\Lambda|^{3/2}U(\xi)/(2\delta)},
\eeq
which corresponds to the equilibrium state.
$Z$ is the normalization factor for $\int\phi_0^2d\xi=1$.
The second lowest eigenfunction $\phi_1$ will represent the metastable mode, and
the corresponding eigenvalue will be connected with the inverse of the lifetime of the metastable state,
$\varepsilon^{1/3}\lambda_1\sim 1/\tau$.

{}From the Schr\"odinger equation (\ref{eq:sch}),
\begin{align}
0&=-\frac{1}{2m}\phi_0''+V\phi_0
\label{eq:lowest} \\
\lambda_1\phi_1&=-\frac{1}{2m}\phi_1''+V\phi_1 .
\label{eq:2lowest}
\end{align}
The mass $m$ is $m=|\Lambda|/(2\delta)$.
If we multiply Eq.~(\ref{eq:lowest}) by $\phi_1$ and Eq.~(\ref{eq:2lowest}) 
by $\phi_0$ and subtract the two equations, we obtain
\beq
\lambda_1\phi_0\phi_1=\frac{1}{2m}(\phi_0'\phi_1-\phi_1'\phi_0)'.
\eeq
Integrating this equation from $-\infty$ to the point C (see Fig. \ref{fig:potential2}),
we obtain
\beq
\lambda_1=-\frac{\phi_1'(C)\phi_0(C)}{2m\int_{-\infty}^C\phi_0\phi_1d\xi}
\label{eq:eigen}
\eeq
because $\phi_0'(C)=0$.
Here we consider the {\it metastable wave function} $\phi_{\rm ms}$,
which corresponds to the localized canonical distribution at the valley of the potential.
Hence we assume for $\xi \lesssim C$
\beq
\phi_0\approx u\phi_{\rm ms} \quad {\rm for}\quad \xi\lesssim C,
\eeq
where $u$ is a constant given by
\beq
u^2=\int_{-\infty}^C\phi_0(\xi)^2d\xi\sim Z^{-1}
\rt{\frac{2\pi\delta}{|\Lambda|^{3/2}U''(A)}}e^{-|\Lambda|^{3/2}U(A)/\delta}.
\eeq
Besides we assume that the first excited eigenfunction is also proportional to $\phi_{\rm ms}$ in the range $\xi\lesssim C$,
\beq
\phi_1\approx v\phi_{\rm ms} \quad {\rm for}\quad \xi\lesssim C,
\eeq
because $\phi_1$ is considered to represent the metastable mode.
Under these assumptions, we get
\beq
\int_{-\infty}^C\phi_0\phi_1d\xi =uv\int_{-\infty}^C\phi_{\rm ms}^2d\xi \approx uv.
\label{eq:WKB1}
\eeq
Using the WKB approximation, it is obtained that
\beq
\phi_1'(C)\approx \rt{2mV(C)}\phi_1(C)=\frac{v}{u}\rt{2mV(C)}\phi_0(C).
\label{eq:WKB2}
\eeq
Substituting Eqs. (\ref{eq:WKB1}) and (\ref{eq:WKB2}) into Eq. (\ref{eq:eigen}),
\beq
\lambda_1=\frac{1}{u^2}\rt{\frac{V(C)}{2m}}\phi_0(C)^2.
\eeq
After some calculation, we obtain
\beq
\lambda_1=\frac{1}{2}\rt{\frac{|\Lambda|}{\pi}|U''(A)U''(C)|}e^{-\beta\Delta F}
\eeq
where the free energy barrier $\beta\Delta F$ is $\beta\Delta F=|\Lambda|^{3/2}(U(C)-U(A))/\delta$.
Therefore the lifetime of the metastable state $\tau$, which is equivalent with the relaxation time, is
\beq
\tau \sim 2N^{1/3}\rt{\pi}|U''(A)U''(C)|^{-1/2}|\Lambda|^{-1/2}e^{\beta\Delta F}.
\eeq
Comparing with the relaxation time obtained by Kramers' argument,
Eq.~(\ref{eq:meta_Kramers}),
they agree with each other except for the minor difference in the 
constant prefactor.

%%miya0426
\section{Monte Carlo simulation}
\label{sec:MC}
We also obtained data by performing kinetic Monte Carlo simulations to confirm 
the data obtained by solving the master equation.
In the Monte Carlo simulations, 
each data point is an average over 1000 samples, 
and the error bars are smaller than the symbol size in 
Fig.~\ref{fig:scaleplot}. 

The algorithm of the Monte Carlo simulations is as follows.
We choose a spin at a site $i$ randomly, and update the 
spin with the probability
corresponding to the Glauber model given by Eq.(\ref{eq:Glauber}):
$$\omega_M(\sigma_i\rightarrow -\sigma_i)=
\frac{\exp\left(-\beta\sigma_i(J(M-\sigma_i)/N+H)\right)}
{2\cosh\left(\beta(J(M-\sigma_i)/N+H)\right)}\Delta t ,$$
where $\sigma_i$ is the spin state of the $i$-th spin ($\sigma_i=\pm 1$).
In principle, a
%Apparently, 
small time increment $\Delta t\ll 1$ is necessary to reproduce the
result of the master equation given as a differential equation.
However, we found almost the same result with the different time division,
$\Delta t$=0.01 and 1 for the quantities plotted in Fig.~\ref{fig:scaleplot}.
Hence we obtained the data with $\Delta t=1$.
%This means that 
During one Monte Carlo step we perform single spin flips $N$ times.
%The above procedure is interpreted as 1 Monte Carlo step.
Therefore, the time $t$ is related to the Monte Carlo step $s$ by
$t=s$ because $\Delta t=1$.
The initial condition of each Monte Carlo simulation is set to  
the spinodal magnetization.

%\bibliography{spinodal_ref.bib} %=================================

\end{document}